# Crystal growth, magnetic, and magnetocaloric properties of $J_{eff}$ = 1/2 quantum antiferromagnet $CeCl_3$

Nashra Pistawala[1], Luminita Harnagea[1], Suman Karmakar[2], Rajeev Rawat[2], and Surjeet Singh[1*]

[1]*Department of Physics, Indian Institute of Science Education and Research, Pune 411008*
[2]*UGC-DAE Consortium for Scientific Research, University Campus, Khandwa Road, Indore 452 001*

* surjeet.singh@iiserpune.ac.in

**Abstract**

We report growth of high-quality single crystals of $CeCl_3$ using a modified Bridgman Stockbarger method in an infrared image furnace. The grown crystals are characterized using single-crystal/powder X-ray diffraction, Laue X-ray diffraction, Raman spectroscopy, magnetization, and heat capacity probes. $CeCl_3$ crystallizes with a hexagonal structure ($P6_3/m$) with a very weak trigonal distortion ($P\bar{3}$). The Raman spectrum at 300 K showcases five, clearly resolvable, phonon modes at 106.8 cm$^{-1}$, 181.2 cm$^{-1}$, 189 cm$^{-1}$, 213 cm$^{-1}$, and 219.7 cm$^{-1}$. The magnetic susceptibility along $H \parallel c$ ($\chi^{\parallel}$) and $H \perp c$ ($\chi^{\perp}$) axis is measured as a function of temperature and magnetic field. While $\chi^{\perp}$ exhibits a broad peak in its temperature variation centred around 50 K, $\chi^{\parallel}$, in contrast, is about two orders of magnitude larger in value and shows a monotonic Curie-like increase upon cooling. This anisotropic behavior with qualitatively different temperature dependences of $\chi^{\perp}$ and $\chi^{\parallel}$ is explained using the crystal field theory. The crystal field in $CeCl_3$ splits the J = 5/2 manifold of $Ce^{3+}$ into three Kramer's doublets with $|5/2,\pm 5/2\rangle$ as the ground state, and $|5/2,\pm 1/2\rangle$ at energy $E_1$ = 61 K, and $|5/2,\pm 3/2\rangle$ at $E_2$ = 218 K as the first and second excited states. Accordingly, M(H) at 2 K along $H \perp c$ is small and shows a linear variation, whereas M(H) along $H \parallel c$ saturates readily (easy-axis) to the expected value. In the specific heat, no magnetic ordering could be seen down to 2 K. However, in non-zero fields the low-temperature specific heat changes dramatically, showcasing a peak at 2.5 K under a moderate field of 30 kOe. The weak Ce-Ce exchange, large Ce moment in the crystal field ground state, and significant anisotropy are all ingredients for realizing a high magnetocaloric effect. Indeed, measurements at low temperatures reveal a maximum entropy change of $-\Delta S_{4f} \approx 23 \pm 1\ J\ Kg^{-1}K^{-1}$ near 2.5 K in the field ranging from 50 kOe to 60 kOe. These values are comparable to some of the best Gd-based magnetocaloric materials, signifying the potential of CeCl$_3$ as a magnetic coolant.

# 1 Introduction

Quantum antiferromagnets have garnered enormous attention in the past few years. They exhibit exotic phases at low temperatures including Quantum Spin Liquids (QSLs) with fractional excitations and long-range entanglement [1]. The triangular and Kagome lattice antiferromagnets are ideal platforms to realize the QSL phase. In 2006, Alexie Kitaev proposed an exactly solvable model for anisotropic spin 1/2 on a honeycomb lattice. The ground state of this model is a QSL with Majorana fermions as emergent quasiparticles [2]. Unlike geometrical frustration, where the magnetic frustration arises due to the geometry of the underlying lattice, in the Kitaev model it is the bond-dependent Ising-like interactions that give rise to a strong frustration, resulting in the QSL behavior [3]. The iridates (e.g., Na$_2$IrO$_3$) and the ruthenates (e.g., α-RuCl$_3$) have been extensively studied to realize the Kitaev model [4]. Search for a half-quantization plateau in the thermal Hall effect, which would confirm the existence of Majoranas, is currently one of the burning topics in quantum condensed matter [5,6]. Recently, YbCl$_3$, has also attracted attention as a quantum antiferromagnet with a honeycomb lattice of the Yb$^{3+}$ ions in their Kramer's doublet ground state (J$_{eff}$ = 1/2) [7–9]. YbCl$_3$ belongs to the rare-earth trichloride family with the general formula RCl$_3$ (R = Rare earth element). While the heavier rare-earth members (R = Tb to Lu) of this series crystallize with a monoclinic structure analogous to α-RuCl$_3$, the lighter members (R = La to Gd) crystallize with a hexagonal structure, where the R$^{3+}$ ions arrange on triangular plaquettes that stack-up to form chains of R$^{3+}$ ions running parallel to the *c*-axis [10–12]. Here, we focus on the member CeCl$_3$ of this series, an antiferromagnetic insulator with an electronic band gap of 4.3 eV [13]. Both Yb$^{3+}$ and Ce$^{3+}$ are Kramer's ions but while Yb$^{3+}$ is just one short of a completely filled *f*-shell, Ce$^{3+}$ has only one electron in the *f*-shell; making it a potential quantum magnet. The previous reports on CeCl$_3$ concentrated only on the low-temperature behavior ranging from 0.1 K to 4.2 K, and in magnetic fields up to 12 kOe [14]. The Nuclear Quadrupole Resonance (NQR) on some rare-earth trichlorides and tribromides, including CeCl$_3$, was reported previously over the same temperature range [15]. According to these reports, in CeCl$_3$, Ce$^{3+}$ ions are in an effective spin ½ ground state, which undergoes long-range antiferromagnetic ordering around 0.11 K. The physical properties of CeCl$_3$ show strong sample dependence, which has been attributed to the presence of impurities in the grown crystals [14,16]. Our study is motivated from the recent interest in quantum materials (e.g., α-RuCl$_3$) showing unusual ground states. While CeCl$_3$ is not isostructural with the α-RuCl$_3$, the heavier members of RCl$_3$ series are, and their growth optimization and properties are therefore of significant interest. The recent theoretical studies on CeCl$_3$, in particular, have shown the emergence of optically driven chiral phonon modes in this material which should generate giant effective magnetic fields (~ $10^2$ T) acting on the paramagnetic 4$f$ spins [17]. This requires high-quality crystals where the induced magnetization can be probed experimentally via the inverse Faraday effect. Despite these interesting theoretical findings, the magnetic and thermodynamic properties of CeCl$_3$ remain unexplored above 4.2 K, where the crystal electric field levels of Ce are expected to dominate the magnetic and thermodynamic properties.

In this study, we report the crystal growth, structural, magnetic and thermodynamic characterizations of high-quality single crystals of CeCl$_3$. It should be pointed out that while the older studies furnish adequate details of the precursor purification, the details of the crystal growth method or the structural characterization of CeCl$_3$ are not well documented. In our work, the mm-size, transparent, high-quality crystals are grown by the modified Bridgman-Stockbarger method in an infrared image furnace. The grown crystals are thoroughly characterized using powder and single-crystal X-ray diffraction, and Laue diffraction. The magnetic susceptibility and specific heat are measured from 2 K to 300 K.

We show that the magnetic susceptibility of CeCl$_3$ crystals shows a huge anisotropy that originates from the crystal electric field splitting of the lowest J-multiplet of Ce$^{3+}$. The susceptibility along the two orientations is fitted using the crystal electric field analysis. The Ce-Ce exchange coupling is estimated to be weak (~ 0.1 K). The low-temperature specific heat changes substantially in the presence of an applied magnetic field, showcasing a peak at 2.5 K under an applied magnetic field of 30 kOe. The zero-field *magnetic* specific heat (C$_{4f}$) is fitted using the crystal field splitting inferred from the magnetic susceptibility data. Given the large magnetic anisotropy and a sizeable ground state magnetic moment, the magnetocaloric assessment was done, which gave a high maximum-magnetic-entropy-change of $-\Delta S_{4f} = 23 \pm 1$ J Kg$^{-1}$K$^{-1}$ near 2 K for $\Delta H = 6\ T$.

## 2 Experimental Methods

Single crystals of anhydrous $CeCl_3$ are grown from a stoichiometric melt using the two different methods: the static Bridgman method and traditional Bridgman-Stockbarger method in an image furnace. The details of the crystal growth methods and experiments are given in Sec. III.A. The single crystal specimens for various experiments were cut inside an Ar-filled glove-box to prevent decomposition upon exposure with the atmospheric air. For this purpose, the crystals were embedded in a thin-layer of mounting wax (used for sample cutting or polishing) before taking them out of the glove-box. At the end of the experiment, the wax layer was dissolved inside the glove-box using dichloromethane ($CH_2Cl_2$), which is both water-repellent and anhydrous. Single crystal X-ray diffraction was carried out using a Bruker Smart Apex Duo diffractometer at 100 K using Mo K$\alpha$ radiation ($\lambda$ = 0.71073 Å). The total exposure time was 2.01 h. The frames were integrated using the Bruker SAINT software package using a narrow frame algorithm. Data were corrected for absorption effects using the multiscan method (SADABS). The structure has been solved and refined using the Bruker SHELXTL software package. The grown crystals were oriented using a Laue camera (Photonic Science, UK) in backscattering geometry using Tungsten as a source material ($\lambda$ = 0.35 Å - 2.5 Å, accelerating voltage 30 kV, Tube current 0.3 mA). The Laue pattern was analyzed using Orient Express 3.4 (V 3.3) software package. Raman spectra were collected at room temperature in backscattering configuration using a Horiba Jobin-Yvon LabRAM HR spectrometer equipped with liquid nitrogen cooled Charge-Coupled Detector (CCD) and laser of 532 nm as a source of excitation. The excitation was maintained at 25% of the maximum power, and the accumulation time for each spectrum was 30 s with 25 iterations each time to get better resolution and intensity of the Raman modes. The $CeCl_3$ crystal was coated with a thin layer of Apiezon N grease to prevent degradation due to its extremely hygroscopic nature. The Raman signal from the Apiezon N was measured separately and subtracted from the total signal to eliminate the background. The Specific-heat measurements were done using the relaxation method in a Physical Property Measurement System (PPMS), Quantum Design, USA. A small piece of crystal, was cut into a 2 mm by 2 mm piece, weighing around 10 mg, mounted on the heat capacity sample holder using a low-temperature Apiezon N grease. The addenda (heat capacity of the sample holder and Apiezon N grease) was measured before loading the sample. Sample degradation from moisture effect is minimized by handling, weighing and cutting of $CeCl_3$ crystal inside the glove box. The magnetic susceptibility was measured at the UGC-DAE CSR, Indore center, Indore, using a VSM probe option in a Physics Property Measurement System (PPMS), Quantum design, USA.

## 3 Results and Discussion

### 3.1 Specific difficulties in the crystal growth of CeCl$_3$

The anhydrous rare-earth trichlorides (RCl$_3$) in general and CeCl$_3$ in particular are highly sensitive to the presence of moisture in the air, reacting readily to form RCl$_3$.xH$_2$O (x ≈ 6), especially during a humid day when this reaction happens over a time scale of few minutes. It is therefore difficult to grow large single crystals of rare-earth trichlorides from a melt as the adsorbed moisture, when not carefully removed, reacts with RCl$_3$ at elevated temperatures leading to a formation of various oxychlorides (e.g., ROCl) and oxides which makes the melt hazy. The grown crystal in such cases also has a hazy appearance with large number of cracks. It is, therefore, necessary to remove the absorbed moisture by dehydrating the hexahydrated RCl$_3$.xH$_2$O under vacuum in the presence of ammonium chloride or halides of carbon or sulphur. These methods have their own drawback as they tend to introduce undesirable impurities (rare-earth carbides or sulfides) [18].

The crystals of rare-earth trichlorides can be grown using the chemical vapor transport method employing AlCl$_3$ as the transporting agent [19]. In this reaction, the rare-earth (R) oxides are first reacted with AlCl$_3$ to form corresponding anhydrous rare-earth trichlorides. These are then further reacted with an excess AlCl$_3$ through a reversible reaction to form gaseous complexes, including RAl$_3$Cl$_{12}$ and RAl$_4$Cl$_{15}$. These gaseous complexes, having a significantly high vapor pressure, are transported to the colder end of the tube, where mm-sized transparent crystals of RCl$_3$ can be extracted. This method is, however, complex and the resulting crystals are small and not of very high quality. Mroczkowski et al. reported crystal growth of EuCl$_3$ using the vertical Bridgman method in the presence of Cl$_2$ gas under high pressure [20]. They purified hydrated precursor by placing it in a stream of HCl gas, before subjecting it to crystal growth. A similar method has been cited for the crystal growth of CeCl$_3$ in Ref. [12]. However, as noted above, Ref. [20] uses Cl$_2$ gas under high-pressure to prevent Eu$^{2+}$ from stabilizing. It is therefore not evident if the complex approaches described in previous studies are necessary presently due to the availability of high-quality precursor materials. We, therefore, used a relatively simpler approach to grown high-quality single crystals of anhydrous CeCl$_3$. Our method involves a simple purifying step followed by crystal growth from a stoichiometric melt. We explored two slightly different methods of crystal growth: the static Bridgman method and traditional Bridgman-Stockbarger method in an image furnace.

The anhydrous $CeCl_3$ powder (Alfa Aesar 99.9%) was stored and handled in an argon filled glove box where $O_2$ and $H_2O$ level is maintained at less than 0.1 ppm at all times. The as-purchased $CeCl_3$ powder was heat-treated under a dynamic vacuum. For this, the $CeCl_3$ powder is loaded in a quartz ampoule that was preheated overnight at 1000°C to remove any adsorbed moisture on the walls of the ampoule. This ampoule was then transferred into the glove-box at 200°C. The mouth of the ampoule was temporarily sealed using a bottle-cork to prevent any moisture from entering it. Inside the glove-box, the as purchased $CeCl_3$ powder was poured into the ampoule. The loaded ampoule was then removed from the glove-box and connected to a turbomolecular pump, while its other end, where the $CeCl_3$ powder is located, is gradually heated in a furnace at 230 °C. After heating at this temperature for 24 h under the dynamic vacuum, the ampoule was allowed to cool down to room temperature before flame-sealing under dynamic vacuum (~ $10^{-5}$ mbar). Up to here the procedure is common for both methods.

*Static Bridgman*: In the case of static Bridgman method, the sealed ampoule was placed in a vertical tubular furnace under a temperature gradient. The furnace was heated to 870 °C, which is higher than the melting point of anhydrous $CeCl_3$ (817 °C), at a rate of 50 °C/h and allowed to dwell at this temperature for 12 h. After this, the furnace was slowly cooled to 750 °C at a rate of 0.3 °C/h, and finally cooled down to room temperature at a rate of 50 °C/h. Shiny transparent crystals measuring up to a few mm in size were extracted from the ampoule.

*Bridgman-Stockbarger*: In this case, the size of the quartz ampoule used was chosen to be around 4-5 mm ID and 8 mm OD. In the first step, the ampoule was placed in a muffle furnace and heated up to 850 °C to obtain a premelted, highly dense ingot about 15-20 mm long, which is subsequently subjected to the crystal growth in an infrared image furnace. For this, the quartz ampoule containing the premelted ingot was loaded in a four-mirror image furnace by suspending it from the upper shaft of the furnace (see, Fig. 1a). The shaft was then lowered until the lower end of the ampoule reached the center of the furnace (the common foci of the four ellipsoidal reflectors), where the light rays converge to form the molten-zone during the floating zone experiment [21]. The lamp power was then gradually raised until CeCl$_3$ started melting. After achieving a homogenous melt, the ampoule was allowed to travel downwards into the region of steep vertical temperature gradient. The lamps used were 1 KW each with the zone-region of about 8-10 mm length. In a mirror furnace with 1 KW lamps, the vertical temperature gradient outside the molten zone is typically several hundred °C/cm. In order to optimize the growth parameters for obtaining crack-free, high-quality crystals, several growth experiments were conducted at various traveling speeds, varying from 1 mm h$^{-1}$ to 0.2 mm h$^{-1}$. At 1 mm h$^{-1}$, the crystal boule developed numerous cracks yielding very small, irregular shaped, crystal pieces. At slower growth speeds the results improved, but even at the slowest growth speed of 0.2 mm h$^{-1}$ the cracks could not be avoided completely. In this case, however, the cracks were fewer, and hence large crystal pieces, several mm long by several mm across, could be obtained. The crystals were cut using a low-speed saw inside the glove box. While cutting, small rectangular crystal pieces cleaved-off the crystal boule. These crystal pieces are fully transparent with atomically flat facets as shown in Fig. 1(b). The flat surfaces of the semicircular piece were dry polished using a silicon-carbide paper of grit-size 1200. The slight haziness seen is due to imperfect surface polishing and not because the crystal piece is hazy inside.

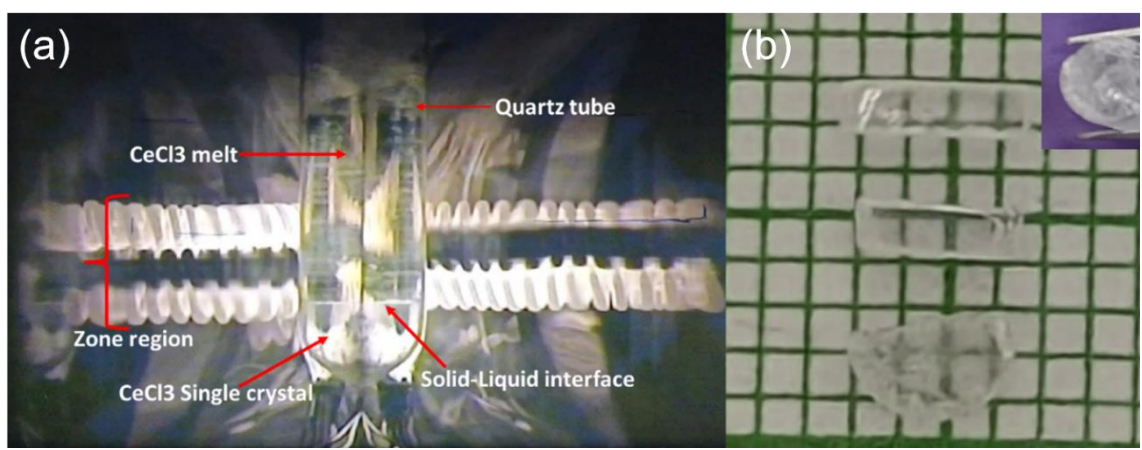

**Fig. 1.** (a) Crystal growth of CeCl$_3$ using a four-mirror image furnace equipped with 1 kW halogen lamps. The lower part below the solid-liquid interface is the CeCl$_3$ crystal being grown. (b) Images of the crystals pieces obtained after cutting the crystal boule shown in the inset.

Between the two growth methods, the crystals obtained using the image furnace are of higher quality. They are fully transparent and hence one can say are essentially defect-free. The presence of defects or impurities (typically oxychlorides and oxides mentioned above) in the crystal leads to hazy or milky appearance. Analysis of impurities in the crystals is summarized in Fig. S1. (see supplementary text). The X-ray diffraction pattern, recorded in the Bragg-Brentano geometry is shown in Fig. 2(a) top, on a flat crystal specimen, such as one shown in the inset. The only reflections seen are those corresponding to the *bc*-plane either in the trigonal or hexagonal symmetries. This shows that the specimen being investigated is an oriented single crystal of high-quality. The orientation of the surface was further confirmed using the X-ray Laue diffraction (see Fig. 2(b)), where sharp Laue spots were observed, which suggest that the grown crystals are of high quality.

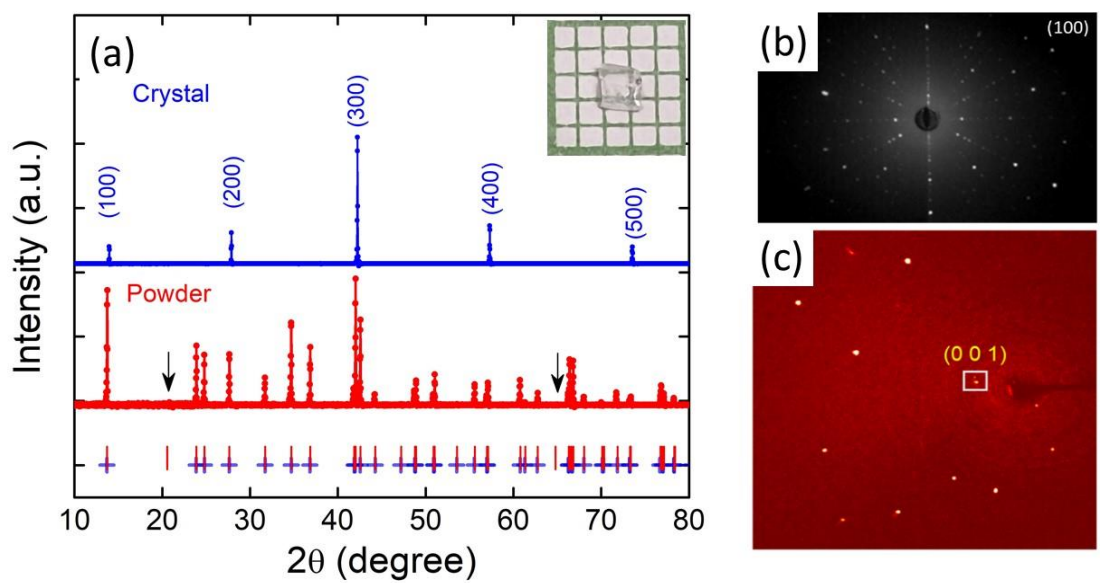

**Fig. 2.** (a) The x-ray diffraction pattern of a single crystal specimen of CeCl$_3$ in the Bragg-Brentano geometry (blue); The powder x-ray diffraction pattern of CeCl$_3$ obtained by crushing a small crystal piece (red); The calculated Bragg positions for hexagonal (+) and trigonal (|) symmetries. The arrows indicate the positions of (0 0 1) and (0 0 3) peaks in the trigonal symmetry. (b) The x-ray Laue diffraction pattern, and (c) A raw frame during the single crystal x-ray diffraction data, where the (0 0 1) spot is marked. Inset in (a) shows a representative crystal specimen used in the x-ray diffraction experiments.

### 3.2 Single crystal X-ray diffraction

A small crystal specimen was selected for single crystal X-ray diffraction (SCXRD). A detailed examination of the collected diffraction data was carried out to establish the crystal structure of CeCl$_3$. The data was collected at T = 100 K. The SCXRD data refinement was done using two different structural models, namely, hexagonal and trigonal. As far as diffraction is concerned, the essential difference between these structures lies in the presence or absence of (0 0 *l*) reflections, where $l = (2n + 1)$. In the trigonal symmetry, these reflections are allowed, whereas in the hexagonal symmetry, they are forbidden. In the raw frames, weak (0 0 *l*) reflections could be identified, as shown in Fig. 2c (see also Fig. S2 in the Supplementary Material for more raw frames). These reflections are better depicted in Fig. S3, where the reconstructed *h*0*l* reciprocal space planes and three-dimensional plots for the (0 0 *l*) reflections are presented. Here, the weak (0 0 1) and (0 0 3) reflections can be seen unambiguously. This is further corroborated by the structural refinement data. In the trigonal case, all the diffraction spots can be successfully indexed, with no systematic absences observed. Conversely, in the hexagonal model, the (0 0 *l*) reflections remain unindexed. In both cases, we used the observability criterion I > 3σ(I), where σ(I) is the standard deviation in the intensity distribution. The intensity (I) and the corresponding σ(I) values for the unindexed reflections under the hexagonal space group are listed in Table. S2 in the Supplementary Material. A comparison of structural refinement parameters for the two models is outlined in Table. 1 below. The goodness of fit and $R_{int}$ values favour the trigonal structure, but the difference in their values is not overwhelming. This is also reflected in the structural parameters obtained in the two cases. For example, in the trigonal model, the space group turned out to be ~~P-3~~ P$\bar{3}$ (147) and the corresponding Wyckoff sites as: Ce(2*d*) with fractional coordinate $\left(\frac{2}{3}, \frac{1}{3}, z_{Ce}\right)$ and Cl(6g) with fractional coordinates $(x_{Cl}, y_{Cl}, z_{Cl})$. The best-fit values obtained for these coordinates are: $z_{Ce} = 0.25004(8)$ and $x_{Cl} = 0.91350(14)$, $y_{Cl} = 0.61249(15)$, $z_{Cl} = 0.7500(3)$. On the other hand, in the hexagonal case, space group $P6_3/m$ (176), the Wyckoff sites came out to be: Ce(2*c*) with fixed fractional coordinate $\left(\frac{2}{3}, \frac{1}{3}, \frac{3}{4}\right)$ and Cl(6h) with fractional coordinates $\left(x_{Cl}, y_{Cl}, \frac{1}{4}\right)$, where $x_{Cl} = 0.91348(15)$, $y_{Cl} = 0.61256(16)$. The coordinates $z_{Ce}$ and $z_{Cl}$ in the trigonal symmetry relates to the hexagonal case by "1 − z". On the other hand, the *x*- and *y*-coordinates of Cl in the two cases differ, if at all, by a very small amount. Besides this difference, and the minute differences in the bond lengths and bond angles, the two structures nearly overlap. To conclude, though the crystal structure of CeCl$_3$ is trigonal in true sense, in practical terms, it deviates from the hexagonal symmetry only marginally. The summary of crystallographic data using the trigonal model is shown in Table 2. The data collection and structure refinement parameters are summarized in Table S3 in the Supplementary Material. and the atomic co-ordinates for Ce and Cl along with their isotropic displacement parameters are listed in Table S4 in the Supplementary Material.

Table 1: Summary of refinement parameters for two different models

| Model | | A | B |
|---|---|---|---|
| Crystal system | | Trigonal | Hexagonal |
| Space group | | $P\bar{3}$ (147) | $P6_3/m$ (176) |
| Goodness of fit | | 1.292 | 1.372 |
| $R_{int}$ (%) | | 4.45 | 4.85 |
| Final R indices (%) | $R_1$ | 2.50 | 2.47 |
| | $wR_2$ | 6.54 | 6.04 |
| Number of refined Parameters | | 14 | 10 |

Table 2. Summary of the crystallographic data in the trigonal model

| Chemical formula | **CeCl$_3$** | |
|---|---|---|
| Formula weight | 246.47 g/mol | |
| Temperature | 100 K | |
| Wavelength | 0.71073 Å | |
| Crystal system | trigonal | |
| Space group | P -3 | |
| Unit cell dimensions | a = 7.4242(15) Å | α = 90° |
| | b = 7.4242(15) Å | β = 90° |
| | c = 4.3189(13) Å | γ = 120° |
| Volume | 206.16(10) Å$^3$ | |
| Z | 2 | |
| Density (calculated) | 3.970 g/cm$^3$ | |
| Absorption coefficient | 12.742 mm$^{-1}$ | |
| F(000) | 218 | |

[*] F(000) is structure factor calculated at h = k = l = 0 and indicates the effective number of electrons in the unit cell

The calculated powder X-ray diffraction profile shows zero-intensity for the (0 0 1) and (0 0 3) lines. The non-zero intensity of these peaks requires the *z*-coordinate of Ce in the trigonal structure to differ sufficiently from the value in the hexagonal case; or for the Cl atoms at 6g (trigonal) or 6h (hexagonal) to have substantially different coordinates in the two structures. Since, the differences in the variable coordinates in the two structures are too small, it will be practically impossible to detect the (0 0 1) or (0 0 3) lines in the powder pattern. The measured powder X-ray diffraction pattern shown in Fig. 2(a) confirms this assertion, as the superstructure lines could not be detected beyond uncertainty. The (0 0 1) line may possibly be picked in a longer run, but keeping the sample undecomposed over a longer duration, even in an airtight sample holder, is a challenge. Using synchrotron-based experiments, one can try to overcome this challenge. In previous studies, the structure of $CeCl_3$ was reported to be hexagonal [10–12]. However, as we found, the deviation from the ideal hexagonal structure is so minor that for all practical purposes one can consider the structure of $CeCl_3$ to be hexagonal.

In both the structures, each $Ce^{3+}$ ion is nine-fold coordinated by $Cl^-$ ions, as shown in Fig. 3(c). Of these, three $Cl^-$ are coplanar with the central $Ce^{3+}$ ion. These are labelled 1, 2, and 3 in Fig. 3(c). The other six $Cl^-$ ions are located above and below the central $Ce^{3+}$ ion. These are labelled from 4, 5, and 6 (above), and 7, 8, and 9 (below) in Fig. 3(c). In the *ac*-plane, shown in Fig. 3(a), the $Ce^{3+}$ ions form a zig-zag chain running parallel to the *c*-axis. The bond distances for the trigonal case are also shown in Fig. 3. The Ce-ions marked with a yellow border are located at a distance of 4.319 Å from the central Ce ion (blue) along the *c*-axis. This is the nearest-neighbor (*nn*) Ce-Ce distance. The three second-nearest neighbors (*nnn*) are at a distance of 4.799 Å. They are shown with a pink border for easy identification. The third nearest neighbor (*nnnn*) distance is 7.424 Å. The third neighbors are shown with a red border, as in Fig. 3(b). The polyhedral around the $Ce^{3+}$ are edge-shared, forming a ring in the *ab*-plane as shown in Fig. 3(b). The magnetic exchange path between any two nearest neighbor Ce ions is via three Ce–Cl–Ce pathways along the *c*-axis, where the Ce–Cl–Ce bond angle is 95.4°, as shown in Fig. 3(d). The *nnn* interaction is mediated via two Ce–Cl–Ce pathways with bond angle Ce–Cl–Ce of 110.1°, as shown in Fig. 3(e).

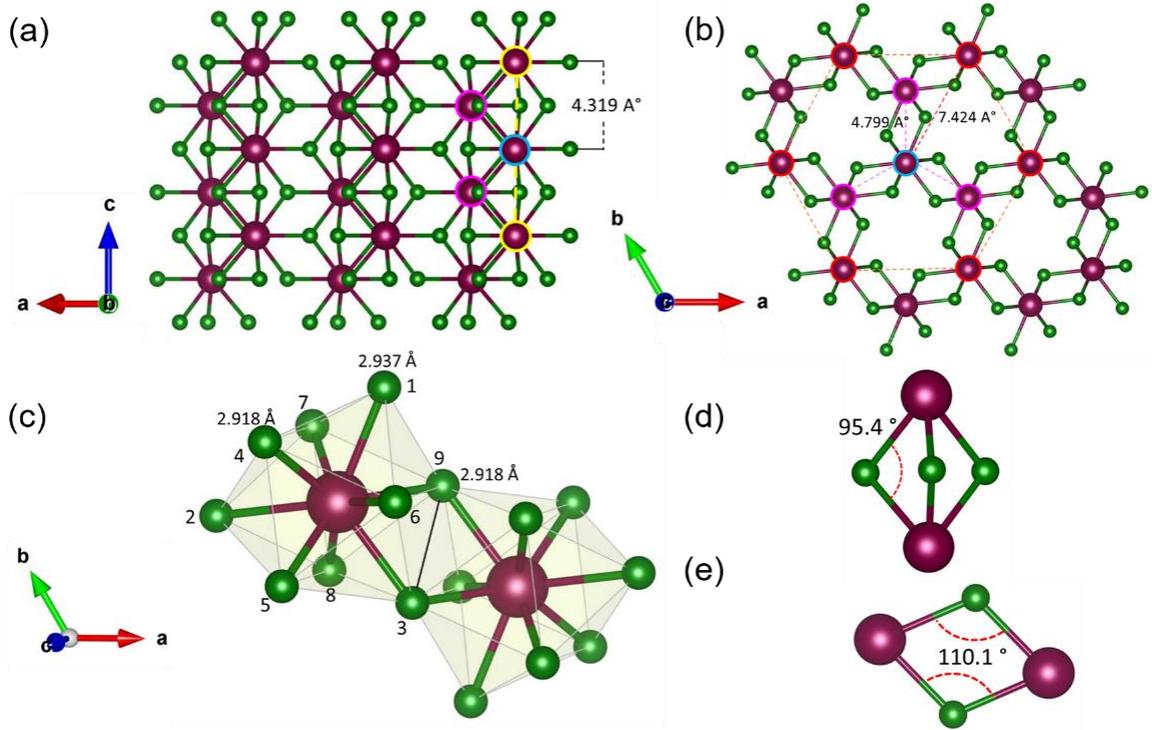

**Fig. 3**. Crystal structure of $CeCl_3$ obtained using VESTA [22]. The brown spheres represent Ce and green spheres (smaller in size) represent Cl. (a) structure as viewed along the *b*-axis. The nearest neighbors around the central Ce (blue border) are shown with yellow borders ($d_{nn}$ = 4.319 Å); (b) View along the *c*-axis showing the second nearest neighbor (magenta border), and the third nearest neighbor Ce (red border) around the central Ce (yellow border). ; $d_{nnn}$ = 4.799 Å, and $d_{nnnn}$ = 7.424 Å (c) 9-fold co-ordination of $Ce^{3+}$ ions where $Cl^-$ ions numbered 1, 2 and 3 are coplanar with the central $Ce^{3+}$ ion. The $Cl^-$ ions numbered 4, 5, and 6 lie in the plane above, and 7, 8, and 9 in the plane below the central plane formed by 1, 2, and 3. These planes are perpendicular to a *c*-axis; (d) Ce-Cl-Ce nearest neighbor exchange pathways via 95.4° angle; (e) second nearest neighbor exchange pathway via 110.1° angle.

### 3.3 Raman Spectrum

Fig. 4 show the Raman spectra of a high-quality single crystal of $CeCl_3$. The observed spectrum is satisfactorily fitted using *five* Lorentzian line shapes. For simplicity, the modes are labelled from $R_1$ to $R_5$ and their positions are: 106.8 cm$^{-1}$, 181.2 cm$^{-1}$, 189 cm$^{-1}$, 213 cm$^{-1}$, and 219.7 cm$^{-1}$. In an older study, $CeCl_3$ is reported to show six Raman modes at $E_{2g}$ (106 cm$^{-1}$), $A_g$ (176 cm$^{-1}$), $E_{2g}$ (180 cm$^{-1}$), $E_{1g}$ (193 cm$^{-1}$), $A_g$ (216 cm$^{-1}$), and $E_{2g}$ (218 cm$^{-1}$) [23]. The comparison with our data suggests the presence of an extra peak at 176 cm$^{-1}$ in the spectra reported in Ref. [21]. Incidentally, in the isostructural $LaCl_3$ only five Raman modes are reported at 108 cm$^{-1}$, 179 cm$^{-1}$, 186 cm$^{-1}$, 210 cm$^{-1}$, and 217 cm$^{-1}$ [24]. The extra mode in $CeCl_3$, as seen in Ref. [21], can be due to the presence of local defects or impurities. Although the polarization dependence of the Raman active mode centered at 179 cm$^{-1}$ was previously reported [23], a more careful polarization dependent study of the entire spectrum is lacking to understand the symmetry ($A_g$ or $E_g$) of all the modes present in $CeCl_3$. Also, a temperature and magnetic field dependent study on high-quality crystals is needed to understand the evolution of Raman modes with temperature and magnetic field. Previously, it was reported that the degenerate $E_{1g}$ and $E_{2g}$ Raman modes split under magnetic field into left- and right-handed circular polarization, leading to a chiral behavior [25]. With the availability of high-quality single crystals of $CeCl_3$, such experiments involving temperature and field dependence will be carried out in future.

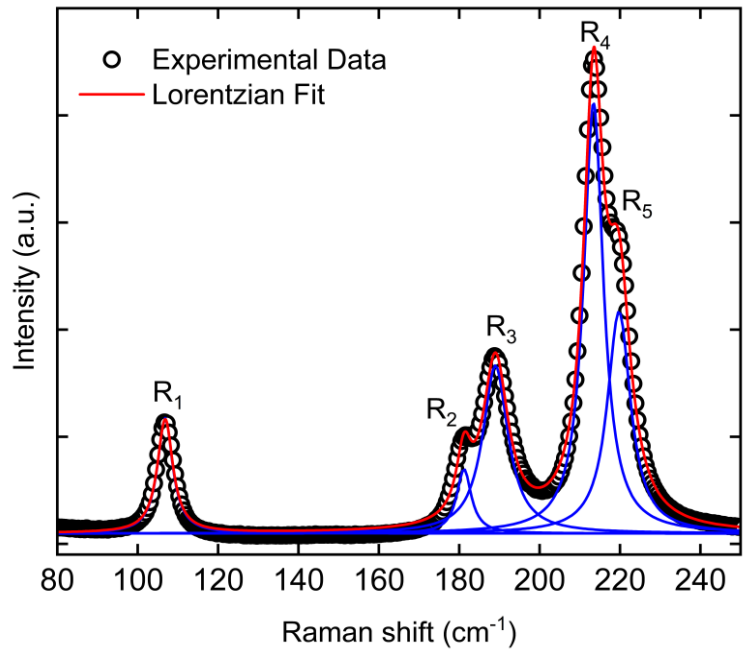

**Fig. 4.** The Raman spectra of $CeCl_3$. The five Raman modes are labelled $R_1, R_2, …, R_5$. The blue curves are individual Lorentzian fits for each mode. The red line through the data point is a total fitted spectrum obtained by adding up the individual Lorentzian. The raw data up to 550 $cm^{-1}$ is shown as Fig. S4 in the Supplementary Material.

### 3.4 Magnetic susceptibility

*Curie-Weiss Analysis*: The magnetic susceptibility of $CeCl_3$ is measured under a magnetic field of 1 kOe for two different orientations, namely $H \parallel c$ and $H \perp c$ axis. Additionally, measurements are also done on a powder specimen of $CeCl_3$, obtained by crushing a small crystal piece. The magnetic susceptibility plots are shown in Fig. 5. The susceptibility measurements are performed in both ZFC and FC modes. The two runs exactly overlaps and hence only ZFC run is shown in Fig. 5. Upon cooling, $\chi^\perp(T)$ (i.e., $\chi(T)$ for $H \perp c$) exhibits a broad peak at 50 K, followed by a sharp increase below 9 K, as shown in Fig. 5. This broad peak is due to the crystalline electric field (CEF) splitting of the J = 5/2 ground state of $Ce^{3+}$. The CEF splitting scheme for $Ce^{3+}$ is discussed in the next section. The Curie-Weiss analysis was performed in the low and high-temperature ranges for both orientations and the powder sample. Due to the presence of a small curvature in $\chi^{-1}$, we used the modified Curie-Weiss expression, $\chi = \chi_0 + C/(T - \theta_{cw})$, where $\chi_0$, $C$, and $\theta_{cw}$ represent the temperature independent van Vleck contribution, Curie constant, and Curie-Weiss temperature, respectively. The symbols $\chi_0^{\parallel}$, $\chi_0^{\perp}$; $C^{\parallel}$, $C^{\perp}$; and $\Theta_{cw}^{\parallel}$, $\Theta_{cw}^{\perp}$ are used to replace $\chi_0$, $C$, and $\theta_{cw}$ in the Curie-Weiss expression for H ∥ $c$ and H ⊥ $c$ orientations, respectively. The value of $\mu_{\text{eff}}$ can be obtained from the Curie constant using the expression $\mu_{\text{eff}} = \sqrt{8C}$. The results are summarized in Table. 3. The fitting for the powder specimen is shown in the inset of Fig. 5, as a representative case.

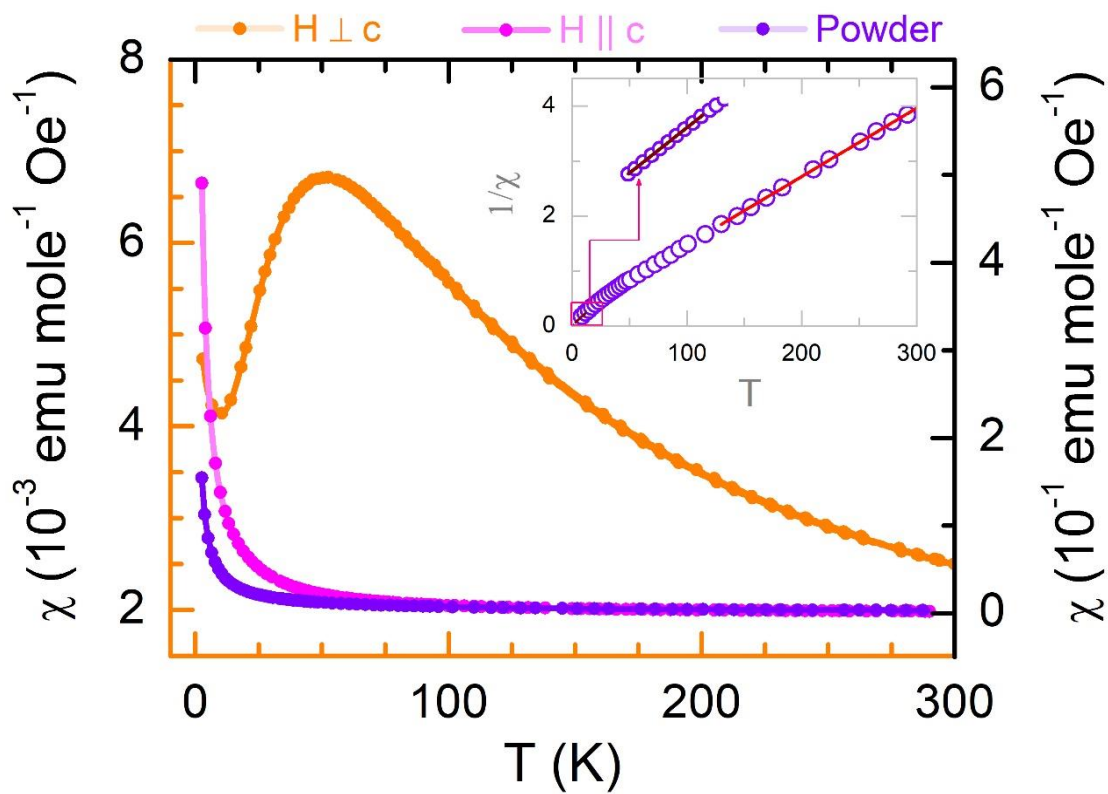

**Fig. 5.** The magnetic susceptibility $\chi$ is plotted as a function of temperature for H ∥ $c$, H ⊥ $c$, and powder specimens of CeCl$_3$. Inset shows the reduced inverse magnetic susceptibility as a function of temperature. The solid lines in the inset represent Curie-Weiss fit (see text for details). In the main panel, the lines are guide to the eye.

The analysis of H ∥ $c$ data yielded the following values of the fitting parameters: $\chi_0^{\parallel} \sim -5.1 \times 10^{-4}$ emu mol$^{-1}$ Oe$^{-1}$, and $C^{\parallel} = 0.81$ emu mol$^{-1}$ Oe$^{-1}$ K giving $\mu^{\parallel}_{\text{eff}} = 2.53(1)\ \mu_B$/Ce. This value of $\mu_{\text{eff}}$ matches closely with the theoretical value of 2.54 $\mu_B$/Ce calculated using J = 5/2 and $g_J = \frac{6}{7}$, where J is the total angular momentum and $g_J$ is the Landé-g factor. The value of $\Theta_{cw}^{\parallel} = 18.7$ K is positive, indicating that the Ce-Ce interaction along the $c$-axis is ferromagnetic. The large value of $\Theta_{cw}^{\parallel}$ compared to the long-range antiferromagnetic ordering temperature ($T_N \sim 0.1$ K [12]) is simply an artifact arising due to the large crystal field splitting. The low temperature fitting is performed in the range 3 K < T < 10 K, yielding $C^{\parallel} = 1.47$ emu mol$^{-1}$ Oe$^{-1}$ K or an effective magnetic moment of 3.42 $\mu_B$/Ce. As we show below, at low temperatures, only the lowest Kramer's doublet $|5/2, \pm 5/2\rangle$ contributes to the magnetic moment. We can therefore estimate $g^{\parallel}$ using the expression $\mu^{\parallel}_{eff} = g^{\parallel}_J \sqrt{J(J+1)}$ by taking J = 1/2 and the experimentally obtained value of 3.43 $\mu_B$/Ce for $\mu^{\parallel}_{eff}$. This gives a value of $g^{\parallel} \sim 4$, which agrees fairly nicely with the value reported previously using the electron spin resonance (ESR) study [26]. The value of $\Theta_{cw}^{\parallel} = 0.017$ K obtained from the low temperature fit still positive, but its value has now been reduced to below 1 K and is closer to the actual ordering temperature.

Similar analysis for $\chi^{\perp}(T)$ yields $\chi_0^{\perp} \sim 2.7 \times 10^{-4}$ emu mol$^{-1}$ Oe$^{-1}$; $C^{\perp} = 0.75$ emu mol$^{-1}$ Oe$^{-1}$, giving $\mu_{eff}^{\perp} = 2.46$ $\mu_B$/Ce. This value of the effective moment is slightly reduced compared to the theoretical value. The value of $\Theta_{cw}^{\perp}$ turned out to be -32.7 K. The negative sign of $\Theta_{cw}^{\perp}$ indicates the interaction between the Ce moments is antiferromagnetic in the $ab$-plane. As above, the enhanced magnitude of $\Theta_{cw}^{\perp}$ is due to the crystal field splitting. From a similar low-temperature fit (see, Table 3), we estimate $g^{\perp}$ to be ~0.17, in good agreement with the ESR value of 0.17.

In samples with large magnetic anisotropy, it is often more insightful to analyze the powder averaged data. The powder susceptibility shown in Fig. 5, agrees well with the calculated values obtained using equation $\chi_{avg} = \frac{2}{3}\chi^{\perp} + \frac{1}{3}\chi^{\parallel}$. The Curie-Weiss analysis similar to the one above yielded the values of fitting parameters consistent with the single crystal values, as shown in Table 3. The negative sign of the Weiss temperature for the powder sample indicates that the average Ce-Ce interaction in CeCl$_3$ is antiferromagnetic in nature. The low-temperature fit gives a value of $\theta_{cw} = -0.12$ K which is negative and small as expected and is closer to the actual ordering temperature.

**Table 3**: The fitting parameter in the Curie-Weiss fit of the susceptibility data for single crystal and powder sample

|  | **Fitting range** (K) | $\chi_0$ emu mole$^{-1}$ Oe$^{-1}$ | C emu mol$^{-1}$ Oe$^{-1}$ K | $\mu_{eff}$ ($\mu_B$) | $\Theta_{cw}$ (K) |
|---|---|---|---|---|---|
| **H ∥ c** | 150 – 300 | $-5.1 \times 10^{-4}$ | 0.81 | 2.54 | 18.7 |

|  | 3 – 10 | $-2.3 \times 10^{-3}$ | 1.47 | 3.42 | 0.017 |
|---|---|---|---|---|---|
| **H ⊥ c** | 200 – 300 | $2.7 \times 10^{-4}$ | 0.75 | 2.46 | -32.7 |
|  | 3 – 7 | $3.8 \times 10^{-3}$ | 0.003 | 0.15 | -0.39 |
| **Powder** | 130 – 300 | $8.8 \times 10^{-4}$ | 0.81 | 2.54 | - 18.2 |
|  | 3 – 10 | $1.8 \times 10^{-3}$ | 0.47 | 1.95 | -0.12 |

*Crystal field analysis*: The degeneracy of lowest J-multiplet in an isolated $Ce^{3+}$ ion ($S = ½$, $L = 3$ and $J = 5/2$) is $2J + 1 = 6$. In a trigonal crystalline electric field (CEF), this degeneracy is lifted by splitting the ground state into three Kramer's doublets. The CEF Hamiltonian is given by:

$$H_{CEF} = \sum_{m,n} B_m^n O_m^n, \tag{1}$$

where $B_m^n$ are the CEF parameters, and $O_m^n$ are the Stevens operators. Since the trigonal distortion in CeCl$_3$ is at best marginal, for the purpose of CEF analysis we can assume the point group symmetry to be hexagonal. With this simplification, the only non-zero parameters in equation (1) are $B_2^0$ and $B_4^0$ [27,28]. Hence, we rewrite equation (1) as:

$$H_{CEF} = B_2^0 O_2^0 + B_4^0 O_4^0 \tag{2}$$

Now, the magnetic susceptibility can be calculated from the van Vleck formula [29]:

$$\chi_i = \frac{2N_A g_J^2 \mu_B^2}{Z} \left[ \sum_n \beta |\langle J_{i,n} \rangle|^2 e^{-\beta E_n} + 2 \sum_{m \neq n} |\langle m|J_{i,n}|n \rangle|^2 \left( \frac{(e^{-\beta E_m} - e^{-\beta E_n})}{E_n - E_m} \right) \right], \tag{3}$$

where, $N_A = 6.023 \times 10^{23}$ is the Avogadro's number, $g_J = 6/7 \sim 0.857$, $\mu_B = 0.927 \times 10^{-20}$ erg Oe$^{-1}$, $\beta = 1/k_B T$, $k_B = 1.38 \times 10^{-16}$ erg K$^{-1}$, $Z = \sum_n e^{-\beta E_n}$, and $n, m = 0, 1, 2$. Here, index *i* corresponds to orientation: $i = x$ for H ⊥ $c$, and $i = z$ for H ∥ $c$. As shown in Ref. [29], for the hexagonal symmetry, the eigenstates of Hamiltonian (2) are simply the unmixed or pure states of the form $|5/2, \pm 5/2\rangle, |5/2, \pm 3/2\rangle$, and $|5/2, \pm 1/2\rangle$. We know from the low-temperature magnetic susceptibility analysis that the crystal field split ground state is $|5/2, \pm 5/2\rangle$, which implies that the first excited state is either $|5/2, \pm 3/2\rangle$ or $|5/2, \pm 1/2\rangle$. We tried both these combinations and found that only in the case where $|5/2, \pm 1/2\rangle$ is taken as the first excited state gives the satisfactory result. The expressions for $\chi^\perp$ and $\chi^\parallel$ using $|5/2, \pm 1/2\rangle$ as the first-excited state and $|5/2, \pm 3/2\rangle$ as the highest state are as follows:

$$\chi^\perp = \frac{N_A g_j^2 \mu_B^2}{Z(k_B)} \left[ \frac{9}{2} \frac{e^{-\beta E_1}}{T} + 5 \left( \frac{1-e^{-\beta E_2}}{E_2} \right) + 8 \left( \frac{(e^{-\beta E_2}-e^{-\beta E_1})}{E_1-E_2} \right) \right] \quad (4)$$

$$\chi^\parallel = \frac{N_A g_j^2 \mu_B^2}{2Z(k_B)} \left( \frac{(25+e^{-\beta E_1}+9e^{-\beta E_2})}{T} \right) \quad (5)$$

where, $Z = 2(1 + e^{-\beta E_1} + e^{-\beta E_2})$, and $E_1$ and $E_2$ are the energy eigen values measured with respect to the ground state doublet. The inverse of expressions (4) and (5) are fitted to the experimental $\chi^{-1}(T)$. Fig. 6(a) and 6(b) show the fitting result for $E_1 = 61\ K$ and $E_2 = 218\ K$. The CEF splitting scheme is shown in the inset of Fig 6(b), where $\Psi_1 = |5/2, \pm 5/2\rangle, \Psi_2 = |5/2, \pm 1/2\rangle$, and $\Psi_3 = |5/2, \pm 3/2\rangle$. These values are in good agreement with Schaack et.al who predicted these splitting to be 47 cm$^{-1}$ (~ 68 K) and 116 cm$^{-1}$ (~ 167 K) using magneto-Raman spectroscopy [25].

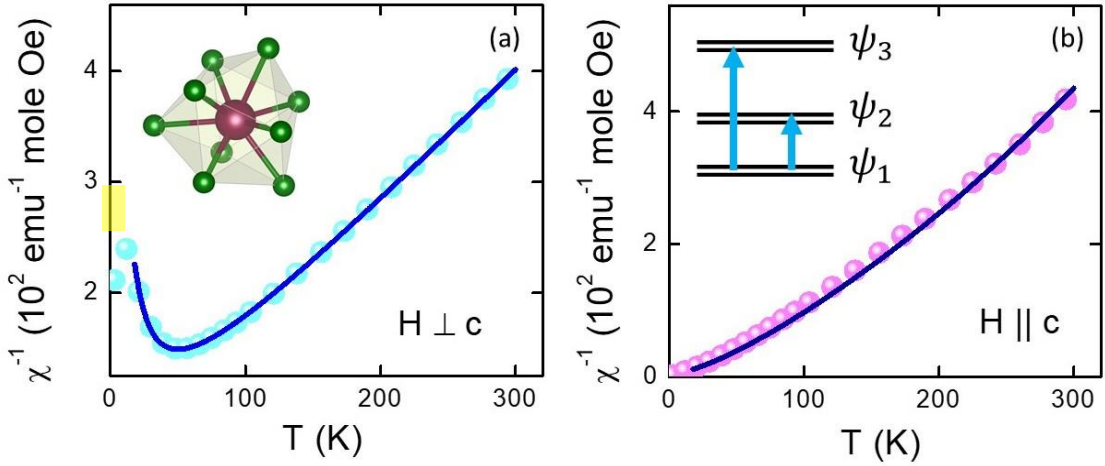

**Fig. 6.** $\chi^{-1}(T)$ is plotted as a function of temperature for the orientations (a) $H \perp c$ (b) $H \parallel c$. The lines through the data points are best fit curves obtained using the crystal electric field analysis (see text for details). The inset in (a) shows the arrangement of ligand ions around the Ce$^{3+}$ ion at the center. The inset in (b) shows the crystal field splitting of the lowest J = 5/2 multiplet of Ce$^{3+}$ in a hexagonal CEF, where $\psi_1 = |\pm 5/2 \rangle; \psi_2 = |\pm 1/2 \rangle, \psi_3 = |\pm 3/2 \rangle$ are the three Kramer's doublets.

### 3.5 Isothermal magnetization

Isothermal magnetization, M(H), plots at 2 K for both orientations ($H \perp c$ and $H \parallel c$) are shown in Fig. 7. A large anisotropy is observed between the orientations. Along the perpendicular orientation, the magnetization is small and linear up to 160 kOe. On the other hand, when the field is applied parallel to *c*-axis, M(H) tends to saturate above a relatively small field of 20 kOe. A saturation magnetization value of $M_s = 2.17\ \mu_B/f.u.$ is close to the saturation moment calculated theoretically, using, $M_s = Jg_\parallel \mu_B \approx 2\ \mu_B$ with $J = \frac{1}{2}$ and $g_\parallel \approx 4$.

The entire range of M(H) data could be nicely fitted using the Brillouin function as shown in Fig. 7.

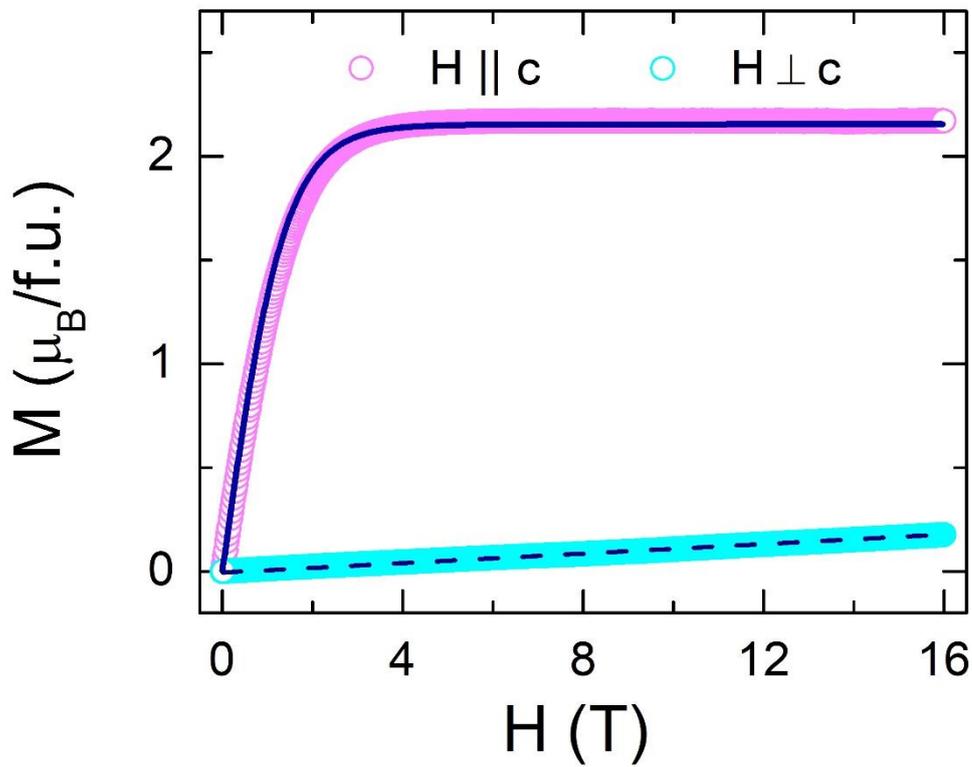

**Fig. 7.** The isothermal magnetization M(H) at 2 K for the orientations $H \perp c$ and $H \parallel c$. The solid line is a fit to the data. The dashed line is a guide to eye (see text for details).

### 3.6 Specific heat

The specific heat of CeCl$_3$ and LaCl$_3$, the latter being the non-magnetic analogue of CeCl$_3$, are measured between 250 K and 2 K as shown in Fig. 8a. The low-temperature data for LaCl$_3$ can be fitted using the equation, $C_p = \gamma T + \beta T^3$, where $\gamma$ and $\beta$ corresponds to electronic and lattice contributions, respectively. LaCl$_3$ being an insulator with an electronic band gap of 5.1 eV [13], the first electronic term can be ignored. The data are replotted as $C_p/T$ vs $T^2$. The variation appears linear (inset in Fig. 8a). Performing a linear-fit yield: $\beta$ = slope = 5.68 ×10$^{-4}$ J mol$^{-1}$ K$^{-4}$. From this, we calculate the Debye temperature $\Theta_D$ = 239 K using the formula $\beta = 12\pi^4 N R/5\Theta_D^3$ where R is the universal gas constant. Note that this value of $\Theta_D$ is not the same as reported by Landau et. al. ($\Theta_D$ = 155 K) [14]. Upon examining their data, we found that the value of $\Theta_D$ obtained by them ignores the factor N (number of atoms per formula unit = 4). At high temperatures, the specific heat rises, towards the Dulong-Petit value of 3NR ($\approx$ 100 J mol$^{-1}$ K$^{-1}$). In order to extract the 4$f$ contribution to the specific heat, we subtracted the lattice part assuming LaCl$_3$ as the lattice template. Being isostructural to CeCl$_3$ and having a comparable molecular weight, LaCl$_3$ is a closest material to model the lattice specific heat of CeCl$_3$. Fig. 8c shows the 4*f* electrons contribution (C$_{4f}$) in zero-field above 2 K. The C$_{4f}$ rises sharply above 2 K, showcasing a broad peak centered around 30 K. This is attributed to the Schottky anomaly associated with the crystal field split Kramer's doublets. C$_{4f}$ is therefore fitted using the three-level Schottky expression, given below:

$$C_{Schottky} = R \frac{1}{T^2} \left\{ \frac{(g_1 g_0 E_1^2 e^{-\beta E_1} + g_2 g_0 E_2^2 e^{-\beta E_2})}{(g_0 + g_1 e^{-\beta E_1} + g_2 e^{-\beta E_2})} + \frac{(g_1 g_2 e^{-\beta(E_1 + E_2)})[E_1(E_1 - E_2) + E_2(E_2 - E_1)]}{(g_0 + g_1 e^{-\beta E_1} + g_2 e^{-\beta E_2})} \right\},$$

where $E_1$ and $E_2$ is the energy of the first and second excited doublets above the ground state, $g_0 = g_1 = g_2 = 2$ is the degeneracy of the states [30]. The best fit to the experimental data is obtained for $E_1 = 67$ K and $E_2 = 200$ K. These values are in close agreement with the values obtained from the susceptibility analysis. While the position of the first excited state is in good agreement with previous report, the energy of the higher lying excited state is somewhat higher than the value of 167 K reported using magneto-Raman spectroscopy [14,25].

Coming now to the field dependence of low-temperature specific heat of our CeCl$_3$ crystal. The magnetic field dependent measurements are summarized in Fig. 8b. The in-field measurements ($H \parallel a$) are superimposed with the data from the previous study [14], where specific heat is reported only below 4.2 K and up to a maximum applied magnetic field of 14 kOe. Our zero-field data overlaps nicely with the zero-field data in Ref. [14]. In zero-field, C$_p$ shows an upturn below 1 K, which is precursory to the long-range ordering of the Ce moments at still lower temperatures at T$_N$ ~ 0.11 K [14]. The magnetic entropy below 2 K ($\approx R\ln2$) corroborates the long-range ordering (*vide infra*). Due to weak exchange coupling between the Ce moments, even a weak applied magnetic field has a strong effect on the specific heat peak, which broadens and shifts toward higher temperatures as the field strength increases, occurring near 2.5 K in an applied magnetic field of 30 kOe. Unfortunately, our attempts to measure the specific heat at higher field turned unsuccessful as the crystal dislodged from the sample platform and shattered into pieces upon cooling below 4.5 K under a field of 50 kOe. This is due to the strong magnetic torque that acts on the specimen due to huge anisotropy and large, but weakly interacting, magnetic moment $\pm 5/2$ in the crystal field split ground state. A slight discrepancy between the in-field data in our study and that taken from Ref. [14] (12 kOe data from Ref. [14] is close to our 20 kOe, indicating a small field off-set), may arise from the actual value of field and the crystal orientation.

We estimated the entropy associated with the 4f electrons, $S_{4f}$, above $T_0 = 2$ K, using the formula: $S_{4f} = \int_{T_0}^{T_1} C_{4f}/T' \, dT'$. As shown in Fig. 8c, the recovered entropy exceeds the Rln2 value near 70 K and continues to rise up to the highest temperature in our measurements. This behavior needs a comment: first, we have not considered a large chunk of $S_{4f}$ buried below 2 K, the range over which the Ce moments order magnetically. If we do this by including the low-temperature specific heat data from previous literature [14], where data are reported between 0.05 K and 4 K, the resulting $S_{4f}$ plot plateau at Rln2 around 2 K, and this plateau stretches up to 10 K. Thus, the first excited state starts contributing to $S_{4f}$ only around 10 K, the temperature above which $C_{4f}$ in Fig. 8c rises sharply. As shown in the Supplementary Material, Fig. S6, the total $S_{4f}$ (i.e., including the low-temperature contribution) overshoots the value $Rln4$ near 70 K. Thus, $S_{4f}$ crossing $Rln2$ near 70 K in Fig. 8c is the contribution sans the Rln2 contribution from the ground state doublet.

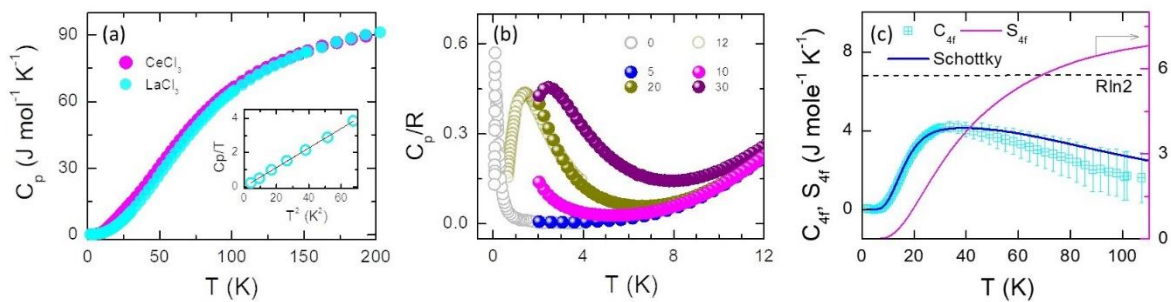

**Fig. 8.** (a) The specific heat ($C_p$) of $CeCl_3$ and its non-magnetic analogue $LaCl_3$ plotted as a function of temperature. Inset shows $C_p/T$ of $LaCl_3$ plotted against $T^2$. The line through the data points is a linear fit to extract the Debye temperature (see text for details). (b) $C_p/R$, where R is the gas constant, plotted as a function of temperature for applied magnetic fields of H = 5, 10, 20, and 30 kOe. Low-temperature zero-field and 12 kOe plots are recreated using data from Landau et al. [14], where $C_p$ is given in the range: T < 4.2 K and H < 14 kOe. (c) $C_{4f}$ and $S_{4f}$ are plotted as a function of temperature. The solid line through the data points ($C_{4f}$) is a three-level Schottky fit using the crystal field splitting estimated from the magnetic susceptibility data (see text for details).

### 3.7 Magnetocaloric Effect

A very small exchange coupling between the Ce-moments and a large moment in the crystal field split ground state of Ce are good indicators that $CeCl_3$ may show promising magnetocaloric properties. In order to estimate the magnetocaloric effect from the indirect method, the isothermal magnetization is plotted as a function of magnetic field for different temperatures to obtain the change in the magnetic entropy of the system when the sample is magnetized up to a maximum applied field of 6T. The measurements were done with applied acting along the crystallographic c-axis. Fig. 9a shows the M(H) isotherms at different temperatures from 2 K to 10 K with temperature interval $\Delta T = 1\ K$. For $\Delta H = 6\ T$, the change in the magnetic entropy is plotted as function of average temperature as shown in Fig. 9b. The maximum value $-\Delta S_m = 23 \pm 1 \text{ J Kg}^{-1}\text{K}^{-1}$ occurs at 2.5 K for $\Delta H = 50 - 60$ kOe. This is a reasonably high value making CeCl3 a potential magnetocaloric as a cryogenic magnetic coolant. However, the highly hygroscopic nature of this material must also be taken into consideration, which may require the coolant material to be permanently sealed in a nonmagnetic but highly electrically conducting capsule. In Table. 4 below we compare the $-\Delta S_m$ for a range of magnetocaloric materials with long-range ordering below 2 K. Among the known Ce-based compounds where magnetocaloric effects has been evaluated, CeCl3 perhaps showcases the highest $-\Delta S_m$ which is in many cases comparable to the Gd-based compounds. This is not surprising given the high J value in the crystal field split ground state of CeCl3. Taking into consideration that the ordering temperature of CeCl3 is an order of magnitude below our measurement temperature, an even higher level and at lower fields is expected if the measurements are extended to lower temperatures.

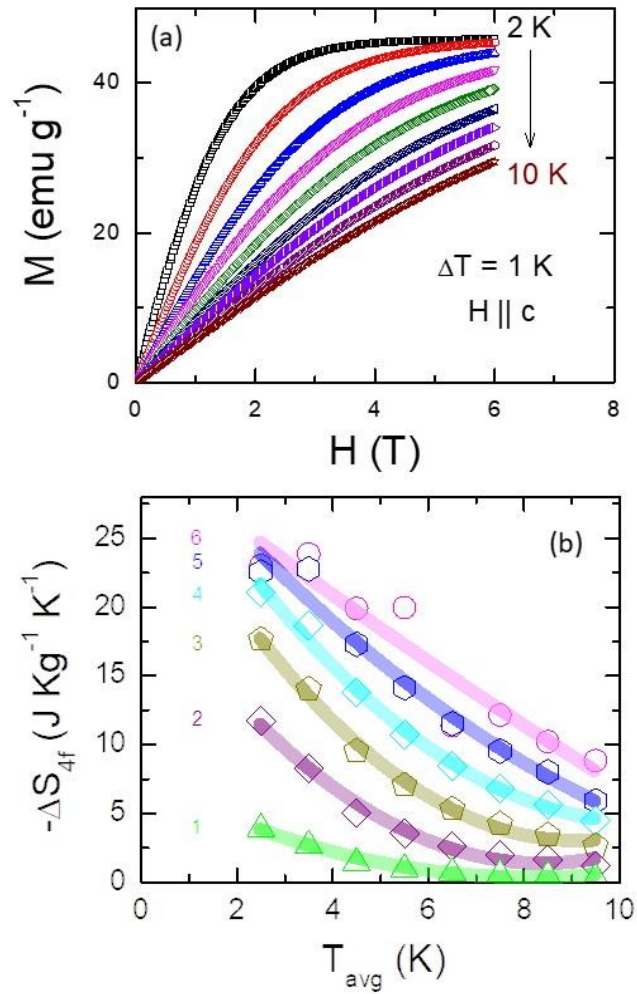

**Fig. 9.** (a) Isothermal magnetization from T = 2 K to 10 K at $T\Delta$ = 1 K interval. The field is applied parallel to the easy-axis (0 0 1) of magnetization. (b) The change in entropy as a measure of the magnetocaloric effect is plotted as a function of $T_{avg}$ (see text for details).

**Table. 4**: A comparison of $-\Delta S_m$ (maximum change in the magnetic entropy) for various previously studied potential cryogenic magnetic coolants at $T_{min}$ (lowest temperature) and $H_{max}$ (maximum magnetic fields)

| Class of material | Compound | $-\Delta S_{max}$ (J Kg$^{-1}$K$^{-1}$) | $T_N/T_C$ (K) | $T_{min}$ (K) | $H_{max}$ (kOe) | Ref. |
|---|---|---|---|---|---|---|
| RCl$_3$ | CeCl$_3$ | 23.0 | 0.11 | 2.5 | 60 | This work |
| Garnets R$_3$Ga$_5$O$_{12}$ | Gd$_3$Ga$_5$O$_{12}$ (GGG) | 35.3 | < 0.025 | 2 | 60 | [31] |
| | Dy$_3$Ga$_5$O$_{12}$ (DGG) | 14.5 | 0.37 | 2 | 60 | [31] |
| | Tb$_3$Ga$_5$O$_{12}$ (TGG) | 11.2 | 0.25 | 4 | 40 | [31] |
| | Nd$_3$Ga$_5$O$_{12}$ (NGG) | 16 | 0.52 | 1 | 50 | [31] |
| R(OH)$_3$ | Er(OH)$_3$ | 26.5 | < 2 | 4 | 50 | [32] |
| RF$_3$ | GdF$_3$ | 67.1 | < 2 | 2 | 50 | [33] |
| | Cd$_{0.9}$Gd$_{0.1}$F$_{2.1}$ | 7.2 | < 2 | 5 | 50 | [34] |
| | Cd$_{0.9}$Tb$_{0.1}$F$_{2.1}$ | 2.7 | < 2 | 5 | 50 | [34] |
| | Cd$_{0.9}$Dy$_{0.1}$F$_{2.1}$ | 0.8 | < 2 | 5 | 50 | [34] |
| Pyrochlores | Gd$_2$Sn$_2$O$_7$ | 32 | 1 | 2 | 90 | [35] |
| | Gd$_2$Ti$_2$O$_7$ | 17.9 | < 2 | 2.5 | 50 | [36] |
| Tripod Kagome | Gd$_3$Mg$_2$Sb$_3$O$_{14}$ | 33 | 1.7 | 2 | 90 | [35] |
| Double perovskite | Ba$_2$GdSbO$_6$ | 24 | < 0.4 | 2 | 70 | [37] |

## 4 Summary and conclusions

High-quality single crystals of CeCl$_3$ are grown using a modified Bridgman and Bridgman-Stockbarger method in an infrared image furnace. Using both methods, mm-sized single crystals were obtained. The crystals obtained using the Bridgman-Stockbarger method were bigger in size, fully transparent, and higher in yield compared to the crystals obtained using the static Bridgman technique. The grown crystals were characterized using single-crystal X-ray diffraction, X-ray diffraction in Bragg-Brentano geometry, X-ray Laue diffraction, and powder X-ray diffraction techniques. While the overall structure is hexagonal, the single-crystal X-ray diffraction show very weak (0 0 *l*) superstructure reflections indicating a slight trigonal distortion. The crystals are highly sensitive to moisture and decompose within few minutes to form white powder (hexahydrate). The low-temperature thermal and magnetic properties of the grown crystals are studied between 2 K and 300 K. The magnetic susceptibility is measured as a function of temperature for two different orientations, namely, $H \parallel c$ and $H \perp c$, in addition to the polycrystalline sample. In $H \perp c$ measurements, a broad and prominent peak, centered around 50 K, is observed. The susceptibility in the $H \parallel c$, on the other hand, in not only two orders of magnitude larger but also shows a monotonic Curie-like increase upon cooling. We analyzed this behavior using the crystal field theory. The ground state J-manifold (J = 5/2) is shown to split into three Kramer's doublets. The eigen states and energies of these doublets are found to be $\psi_1 = |5/2, \pm 5/2\rangle$ (the ground state), $\psi_2 = |5/2, \pm 1/2\rangle$ (the first excite state at $E_1 \approx 61$ K), and $\psi_3 = |5/2, \pm 1/2\rangle$ (the second excited doublet at $E_2 \approx 218$ K). Using these parameters, the susceptibility behavior along the two orientations is successfully described. The crystal field ground state of $Ce^{3+}$ is nearly pure $|\pm 5/2\rangle$, wherein the large Ce moment is constrained to point along the $c$-axis, acting like an Ising system. This is nicely manifested in the isothermal magnetization measured along the two orientations. The M(H) curve is linear and the M(H) values are very small $H \perp c$; whereas, M(H) saturates readily above ~ 30 kOe in the $H \parallel c$ orientation, depicting that $c$-axis is the easy-axis of magnetization.

The specific heat is measured in the temperature range from 2 K to 300 K. No anomalies could be seen at any temperature. The weak upturn at very low temperatures is attributed to the short-range correlations between the Ce moments. The analysis of 4$f$ derived magnetic specific heat, showcasing a Schottky anomaly, is successfully done using the crystal field scheme obtained from the magnetic susceptibility analysis. The low-temperature magnetic field undergo spectacular changes in the presence of a magnetic field. With increasing applied magnetic field, the low-temperature upturn becomes more pronounced and shifts to higher temperatures, showcasing a peak above 2 K in fields as small as 25 kOe.

The weak exchange between the Ce moments, the huge Ising-like anisotropy, and the large magnetic moment are ideal settings to realizing high magnetocaloric effect. Indeed, we found a maximum entropy change of $-\Delta S_m = 23 \pm 1$ $J Kg^{-1} K^{-1}$ at 2.5 K in the field range 50-60 kOe. This value of $-\Delta S_m$ is rather impressive and in the same ballpark as the best results in previous literature on some Gd based compounds. Therefore, CeCl$_3$ is potential magnetocaloric material for cryogenic applications as a magnetic coolant. The Raman spectrum of CeCl$_3$ exhibits five Raman active modes at modes at 106.8 cm$^{-1}$, 181.2 cm$^{-1}$, 189 cm$^{-1}$, 213 cm$^{-1}$, and 219.7 cm$^{-1}$. In future, polarization, temperature, and magnetic field dependent Raman spectroscopy are needed to completely understand the evolution of Raman modes and the correlation between the spin and lattice degrees of freedom. The degenerate infrared active E$_{1u}$ phonon mode will be interesting to study in future as the electric dipole moment associated with this phonon can be resonantly excited by the electric field of the applied laser pulse. This can yield large vibrational amplitudes that can act on the spins through the inverse spin-phonon coupling as predicted in a recent theoretical work [17]. Since, the *nn* and *nnn* exchange interactions are of comparable magnitude, the bond angles Ce-Cl-Ce and bond distances Ce-Ce being of comparable values for the two exchange pathways, further investigations using neutron scattering and muon spin relaxation techniques would be useful in capturing the exact magnetic ground state and excitations of CeCl$_3$.

**Figure captions:**

**Fig. 1.** (a) Crystal growth of CeCl$_3$ using a four-mirror image furnace equipped with 1 kW halogen lamps. The lower part below the solid-liquid interface is the CeCl$_3$ crystal being grown. (b) Images of the crystals pieces obtained after cutting the crystal boule shown in the inset.

**Fig. 2.** (a) The x-ray diffraction pattern of a single crystal specimen of CeCl$_3$ in the Bragg-Brentano geometry (blue); The powder x-ray diffraction pattern of CeCl$_3$ obtained by crushing a small crystal piece (red); The calculated Bragg positions for hexagonal (+) and trigonal (|) symmetries. The arrows indicate the positions of (0 0 1) and (0 0 3) peaks in the trigonal symmetry. (b) The x-ray Laue diffraction pattern, and (c) A raw frame during the single crystal x-ray diffraction data, where the (0 0 1) spot is marked. Inset in (a) shows a representative crystal specimen used in the x-ray diffraction experiments.

**Fig. 3**. Crystal structure of CeCl$_3$ obtained using VESTA [22]. The brown balls represent Ce and smaller green balls represent Cl. (a) View along the *b*-axis (*ac*-plane) indicating nearest neighbor in yellow circles; (b) View along the *c*-axis (*ab*-plane) indicating second nearest neighbor in magenta and third nearest neighbors in red; (c) 9-fold co-ordination of Ce$^{3+}$ ions where Cl$^-$ ions numbered 1, 2 and 3 are coplanar with the central Ce$^{3+}$ ion. The Cl$^-$ ions numbered 4-6 lie in a plane above and 7-9 in a plane below the plane formed by 1-3. These planes are perpendicular to a c-axis; (d) Ce-Cl-Ce nearest neighbor exchange pathways via 95.4° angle; (e) second nearest neighbor exchange pathway via 110.4° angle.

**Fig. 4.** The Raman spectra of CeCl$_3$. The five Raman modes are labelled R$_1$, R$_2$, …, R$_5$. The blue curves are individual Lorentzian fits for each mode. The red line through the data point is a total fitted spectrum obtained by adding up the individual Lorentzian. The raw data up to 550 cm$^{-1}$ is shown in Fig. S4 in the Supplementary Material.

**Fig. 5.** The magnetic susceptibility $\chi$ is plotted as a function of temperature for H ∥ $c$, H ⊥ $c$, and powder specimens of CeCl$_3$. Inset shows the reduced inverse magnetic susceptibility as a function of temperature. The solid lines in the inset represent Curie-Weiss fit (see text for details). In the main panel, the lines are guide to the eye.

**Fig. 6.** $\chi^{-1}(T)$ is plotted as a function of temperature for the orientations (a) $H \perp c$ (b) $H \parallel c$. The lines through the data points are best fit curves obtained using the crystal electric field analysis (see text for details). The inset in (a) shows the arrangement of ligand ions around the $Ce^{3+}$ ion at the center. The inset in (b) shows the crystal field splitting of the lowest J = 5/2 multiplet of $Ce^{3+}$ in a hexagonal CEF, where $\psi_1 = |\pm 5/2>$; $\psi_2 = |\pm 1/2>$, $\psi_3 = |\pm 3/2>$ are the three Kramer's doublets.

**Fig. 7.** The isothermal magnetization M(H) at 2 K for the orientations $H \perp c$ and $H \parallel c$. The solid line is a fit to the data. The dashed line is a guide to eye (see text for details).

**Fig. 8.** ((a) The specific heat ($C_p$) of $CeCl_3$ and its non-magnetic analogue $LaCl_3$ plotted as a function of temperature. Inset shows $C_p/T$ of $LaCl_3$ plotted against $T^2$. The line through the data points is a linear fit to extract the Debye temperature (see text for details). (b) $C_p/R$, where R is the gas constant, plotted as a function of temperature for applied magnetic fields of H = 5, 10, 20, and 30 kOe. Low-temperature zero-field and 12 kOe plots are recreated using data from Landau et al. [14], where $C_p$ is given in the range: T < 4.2 K and H < 14 kOe. (c) $C_{4f}$ and $S_{4f}$ are plotted as a function of temperature. The solid line through the data points ($C_{4f}$) is a three-level Schottky fit using the crystal field splitting estimated from the magnetic susceptibility data (see text for details).

**Fig. 9.** (a) Isothermal magnetization from T = 2 K to 10 K at $\Delta T = 1$ K interval. The field is applied parallel to the easy-axis (0 0 1) of magnetization. (b) The change in entropy as a measure of the magnetocaloric effect is plotted as a function of $T_{avg}$ (see text for details).

# Supplementary Material

**Crystal growth, magnetic, and magnetocaloric properties of $J_{eff}$ = 1/2 quantum antiferromagnet CeCl$_3$**

Nashra Pistawala[1], Luminita Harnagea[1], Suman Karmakar[2], Rajeev Rawat[2] and Surjeet Singh*[1]

[1]Department of Physics, Indian Institute of Science Education and Research, Pune 411008
[2]UGC-DAE Consortium for Scientific Research, University Campus, Khandwa Road, Indore 452 001

* surjeet.singh@iiserpune.ac.in

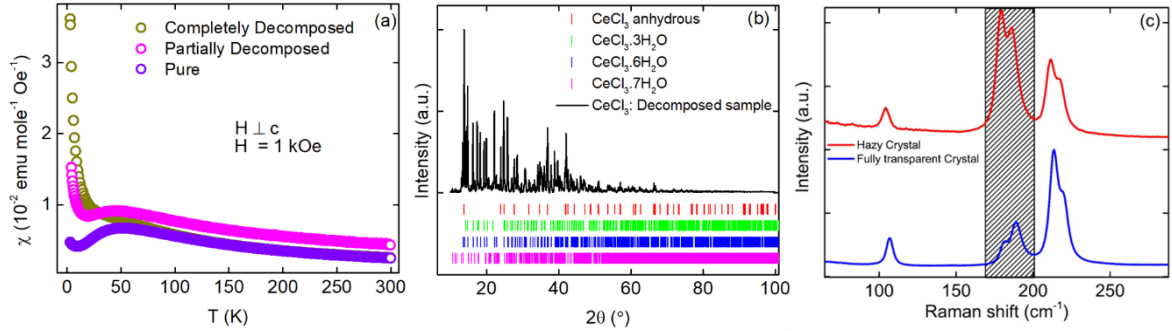

**Fig. S1**. (a) Magnetic susceptibility for the perpendicular orientation (H ⊥ c) for: (i) a fully transparent crystal (labelled as 'Pure' and shown using the purple colour), (ii) a hazy specimen, which is partially decomposed (shown using the magenta colour), (iii) a fully decomposed sample (shown using the dark yellow colour). Note that the decomposition was allowed after the specimen was mounted on the sample holder. Though the decomposed crystal turns into a powder, the orientation of the particles in the powder is expected to remain the same. Note the increase in the Curie-tail upon decomposition. (b) Powder X-ray diffraction of a decomposed specimen showing diffraction peaks due to the formation of CeCl$_3$.xH$_2$O species. (c) Raman spectra of a transparent crystal specimen (blue) and a hazy crystal specimen (red).

Table S1. The comparison of lattice parameter between earlier reports and the present study

| Publication | Space group | Atomic positions | | Lattice parameters (Å) | |
|---|---|---|---|---|---|
| | | Ce | Cl | | |
| W. H. Zachariasen [1] | $P6_3/m$ (176) | 2c | 6h | a = 7.436(4) | c = 4.304(4) |
| D H Templeton [2] | $P6_3/m$ (176) | 2c | 6h | a = 7.450(4) | c = 4.315(2) |
| Present study | $P\bar{3}$ (147) | 2d | 6g | a = 7.4242(15) | c = 4.3189(13) |

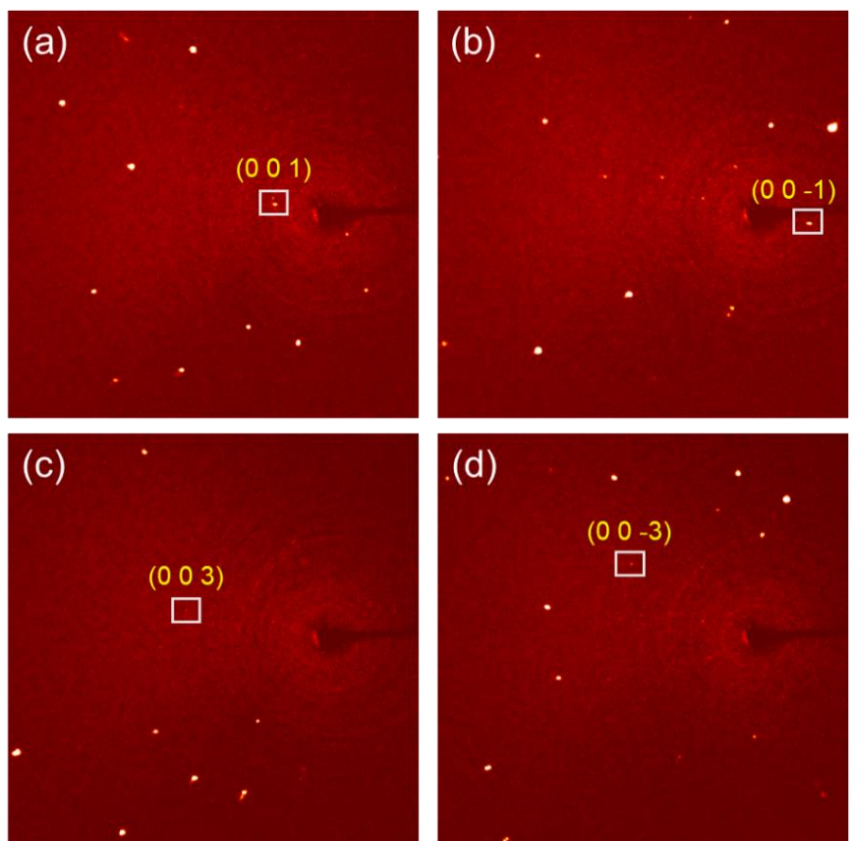

**Fig. S2.** *(0 0 l)* reflections observed in the raw frames of single crystal x-ray diffraction experiment. (a) (0 0 1) (b) (0 0 -1) (c) (0 0 3) (d) (0 0 -3)

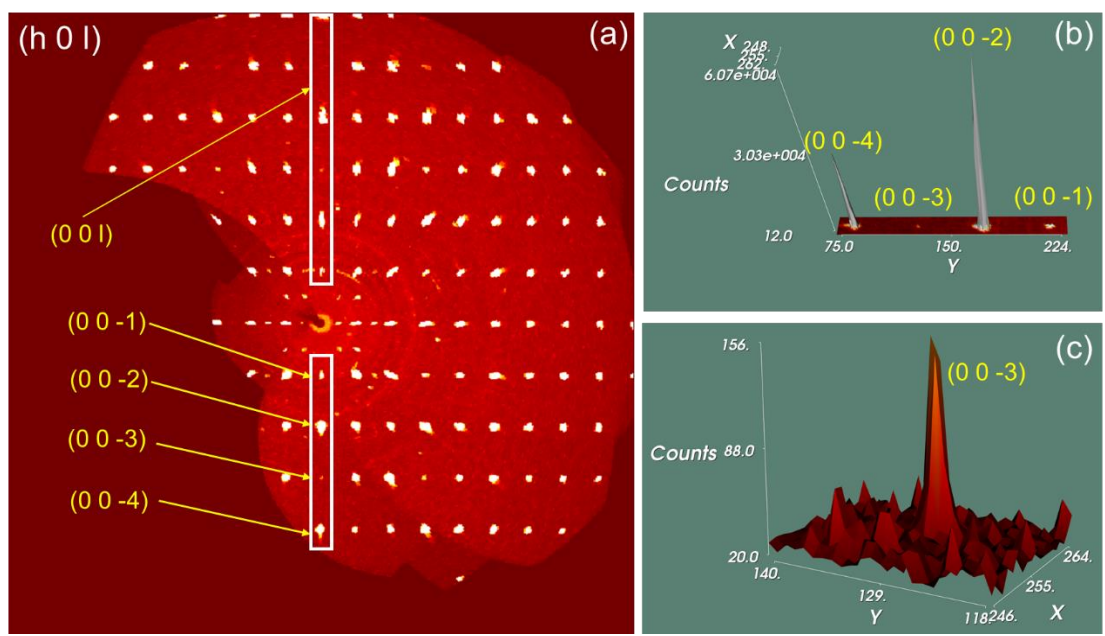

**Fig. S3.** (a) Reconstructed plots of reciprocal space along *(h 0 l)* direction. The X and Y scale are position of spot measured in the units of detector pixel (b) Three-dimensional view of (0 0 l) reflections marked with a rectangle in the lower part of (a). (c) Three-dimensional view of (0 0 -3) reflection.

**Table S2.** Intensity and systematic absent reflections in hexagonal space group

| h k l | I | σ(I) | Why rejected |
|-------|------|------|--------------|
| *0 0 1* | *21.60* | *2.20* | |
| 0 0 1 | 17.10 | 1.90 | Systematically absent |
| 0 0 1 | 21.90 | 2.20 | but $I > 3\sigma(I)$ |
| 0 0 1 | 20.50 | 2.80 | |
| 0 0 3 | 22.50 | 6.80 | |

**Table S3.** Data collection and structure refinement for CeCl$_3$ single crystal in the trigonal model

| **Theta range for data collection** | **3.17 to 29.06°** |
|---|---|
| Index ranges | -8<=h<=10, -9<=k<=10, -5<=l<=5 |
| Reflections collected | 2441 |
| Independent reflections | 374 [R(int) = 0.0445] |
| Structure solution technique | direct methods |
| Structure solution program | XT, VERSION 2018/2 |
| Refinement method | Full-matrix least-squares on F$^2$ |
| Refinement program | SHELXL-2019/1 (Sheldrick, 2019) |
| Function minimized | $\Sigma\ w(F_o^2 - F_c^2)^2$ |
| Data / restraints / parameters | 374 / 0 / 14 |
| Goodness-of-fit on F$^2$ | 1.292 |
| Final R indices | 369 data; I>2σ(I) | R1 = 0.0250, wR2 = 0.0654 |
| | all data | R1 = 0.0257, wR2 = 0.0658 |
| Weighting scheme | w=1/[σ$^2$(F$_o^2$)+(0.0317P)$^2$+0.8350P] where P=(F$_o^2$+2F$_c^2$)/3 |
| Extinction coefficient | 0.0590(60) |
| Largest diff. peak and hole | 1.722 and -1.448 eÅ$^{-3}$ |
| Root mean square (RMS) deviation from mean | 0.380 eÅ$^{-3}$ |

$^£$ F is the structure factor and is proportional to the square root of the intensity of reflections; $^€$ F$_o$ and F$_c$ are the observed and calculated structure factors, respectively; $^§$ R$_1$ is Residual Factor, and wR$_2$ is weighted Residual factor.

**Table S4.** Atomic coordinates and equivalent isotropic atomic displacement parameters (Å$^2$)

| Atoms | x/a | y/b | z/c | U(eq) |
|-------|------|------|------|-------|
| Ce | 0.666667 | 0.333333 | 0.25004(8) | 0.0037(2) |
| Cl | 0.91350(14) | 0.61249(15) | 0.7500(3) | 0.0048(3) |

U(eq) is defined as one third of the trace of the orthogonalized U$_{ij}$ tensor.

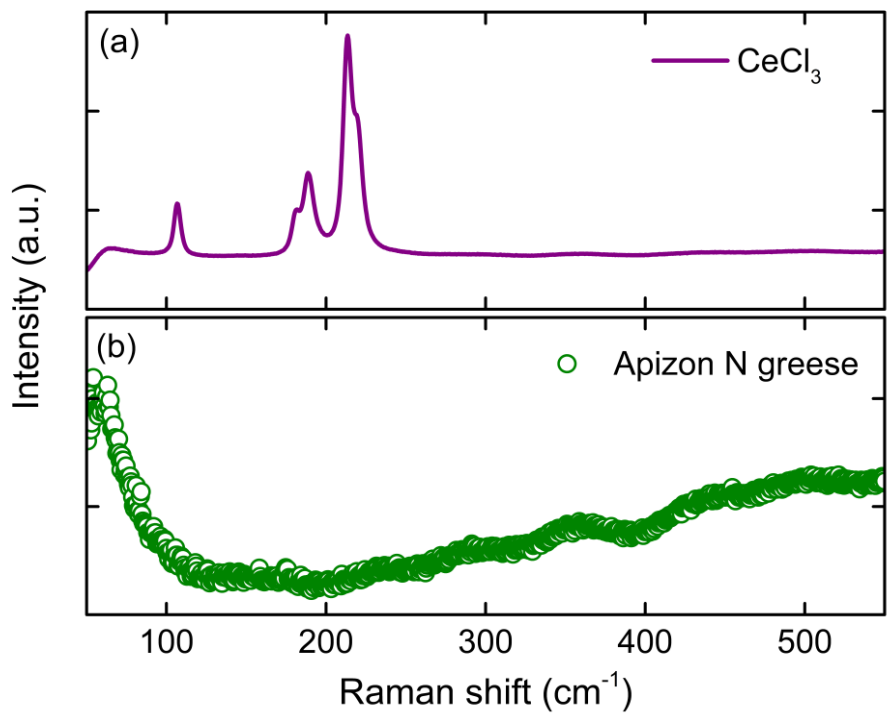

**Fig. S4.** Raman Spectra. (a) CeCl$_3$ Data collected at room temperature. (b) Apiezon N grease data collected at room temperature.

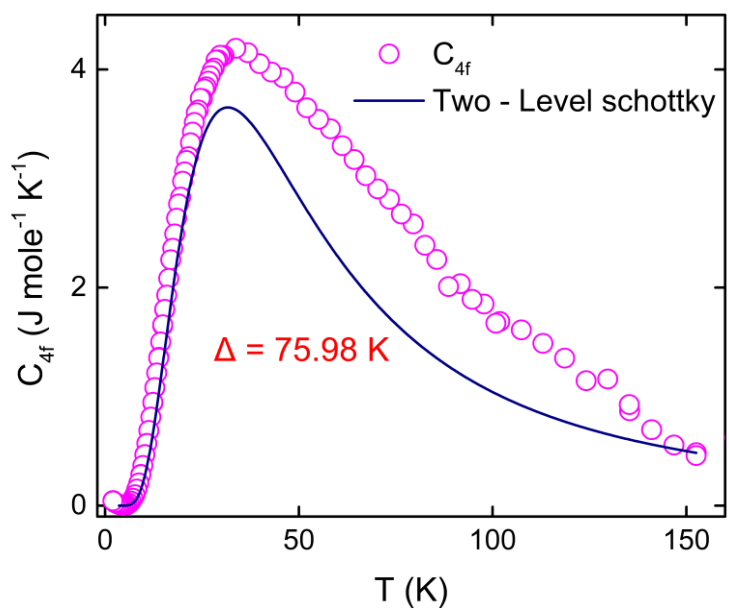

**Fig. S5.** The $C_{4f}$ is fitted using two-level Schottky model. The fitting clearly indicates that two level Schottky equation is not sufficient to model the experimental data and hence $C_{4f}$ involves contribution from the second excited doublet.

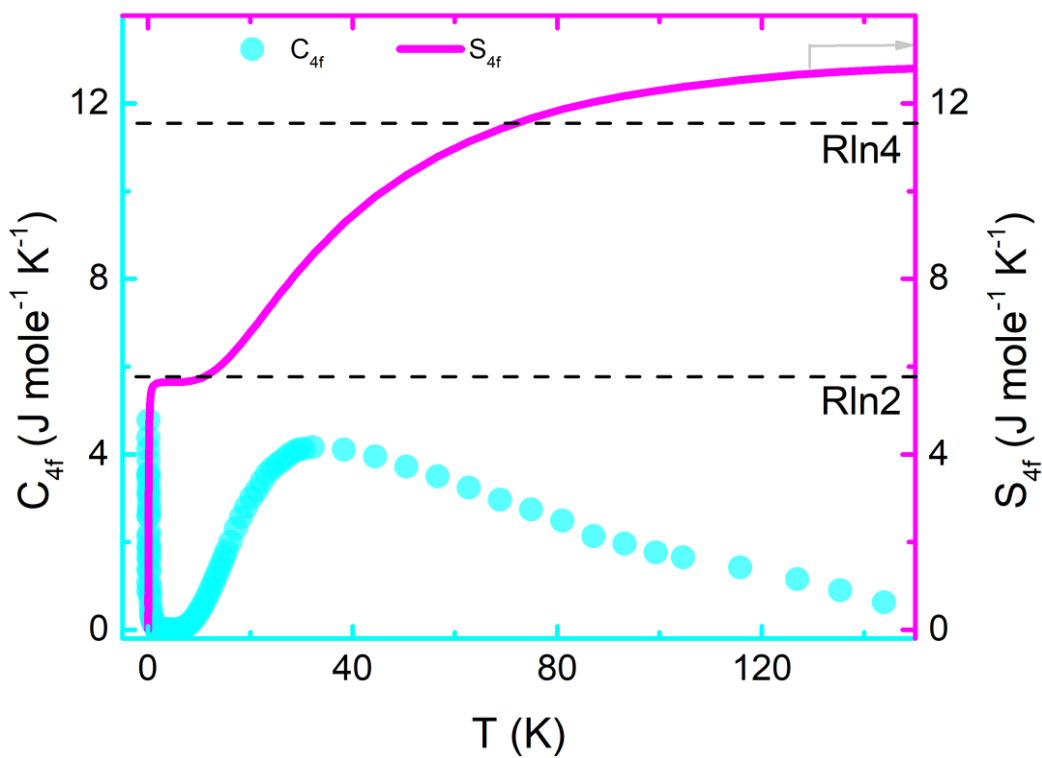

**Fig. S6.** Contribution to specific heat and Entropy due to 4f electrons. The $C_{4f}$ clearly shows a two-peak feature, the low temperature feature is attributed to the ordering of $Ce^{3+}$ ions and the broad peak centered around 30 K is due to CEF splitting of $J = 5/2$ multiplet into three doublets. This is reflected in the Entropy data (see main text for details). The data below 2 K is taken from P. Landau et. al. [3]